\documentclass[epj]{svjour}
\usepackage{graphics}
\begin{document}
\title{Studying the $\omega$-mass in-medium in $\gamma + A\rightarrow \pi^0\gamma +X$ reactions}
\author{J.G.~Messchendorp\inst{1}, A.~Sibirtsev\inst{2,3}, W.~Cassing\inst{2}, V.~Metag\inst{1}, S.~Schadmand\inst{1}}
\titlerunning{Studying the $\omega$-mass in-medium in $\gamma + A\rightarrow \pi^0\gamma +X$ reactions}
\authorrunning{J.G.~Messchendorp et al.}

                     % Do not remove
%
\offprints{}          % Insert a name or remove this line
\institute{II. Physikalisches Institut, Universit\"at Gie{\ss}en,
35392 Gie{\ss}en, Germany \and
Institut f\"ur Theoretische Physik, Universit\"at Gie{\ss}en,
35392 Gie{\ss}en, Germany \and
Institut f\"ur Kernphysik, Forschungszentrum J\"ulich GmbH, 52425 J\"ulich, Germany}
\date{Received: date / Revised version: date}
% The correct dates will be entered by Springer
%
\abstract{Simulations based on a coupled-channel transport model have
been performed to analyze the feasibility to study the in-medium $\omega$ mass exploiting the process
$\gamma+A\rightarrow\pi^0\gamma+X$ for $C$, $Ca$ and $Nb$ nuclei.
The distortions due to final-state
interactions of the $\pi^0$ and background contributions from the reaction
$\gamma+A\rightarrow\pi^0\pi^0+X$ are found to be small in the mass range of
interest ($0.6<M_{\pi^0\gamma}<0.8$~GeV). Furthermore, the effect of
the detector resolution on the $\pi^0\gamma$-mass determination
is discussed.
}
\maketitle
\section{Introduction}
\label{intro}

\begin{sloppypar}
The modification of hadron properties in the nuclear medium is one of
the current topics in hadron and nuclear physics.
Vector-meson properties in the nuclear medium have attracted
considerable attention since their in-medium
spectral function was measured by dilepton spectroscopy ($l^+l^-$)
without distortion due to final-state interactions.
An enhancement in the lepton pair spectrum from heavy-ion
collisions relative to proton-nucleus reactions has been observed
by the CERES~\cite{Agakichiev1,Agakichiev2,Agakichiev3} and
HELIOS-3~\cite{Masera} collaborations at the SPS in
the invariant-mass range from 0.4 -- 0.7~GeV. As proposed in
Refs.~\cite{Brown,Li,Cassing1,Rapp1,Cassing2} an enhancement
can be understood in terms of a modification of the vector-meson
properties in dense and hot nuclear matter. Collisional broadening
of the $\rho$-meson width or dropping of
the $\rho$-mass pole might be responsible for the experimental
observations. The in-medium modifications of vector mesons have
also been discussed in Refs~\cite{Boffi,Bianchi1,Rapp2,Muccifora,Rapp3} in
the context of photo-absorption data on
nuclei~\cite{Bianchi1,Bianchi2}. Recently, it was
suggested~\cite{Falter} that photoproduction data can
provide access to the momentum dependence of the in-medium
$\rho$-meson potential~\cite{Eletsky1,Kondratyuk,Eletsky2}.
\end{sloppypar}

Theoretical studies within different models predict a dropping of
the in-medium $\rho$-meson mass ($m^\ast$) at normal
nuclear-matter density ($\rho_0{=}$0.16 fm$^{-3}$) within the range of
$-240{\le}m^\ast-m_V{\le}-45$~MeV~\cite{Hatsuda1,Asakawa,Koike,Jin,Saito1,Tsushima1,Leupold},
where $m_V$ denotes the bare $\rho$-mass. However, it was argued
in Refs.~\cite{Chanfray,Herrmann,Rapp4,Friman1,Klingl1} that due
to substantial collisional broadening the $\rho$-meson might not
survive as a ``proper'' quasiparticle at large nuclear densities and
only reflects a continuum spectral strength of the isovector
current. Thus, the $\rho$-meson spectral function is difficult to
identify at high baryon density.

\begin{sloppypar}
The situation is better for the $\omega$-meson. Here, the
predictions  for  a modification of the $\omega$-mass (at
normal nuclear-matter density $\rho_o$) are within the
range $-140{\le}m^\ast -m_V{\le}-15$ MeV
\cite{Tsushima1,Klingl1,Friman2,Saito2,Saito3,Klingl2}, while the
$\omega$-meson collisional width is estimated within the range of
20 to 50 MeV~\cite{Klingl1,Friman2,Lykasov}. Thus it is expected
that the in-medium $\omega$-meson survives as a quasiparticle
and can be observed as a structure in the $\omega$-mass spectrum.
In this case, one should observe a significant resonance structure on top
of a continuum.
\end{sloppypar}

The $\phi$ and $J/\Psi$ mesons
should also change their masses slightly at normal nuclear density
with a small collisional width due to the almost
negligible interaction with nucleons
\cite{Koike,Jin,Tsushima1,Klingl1}. However, their in-medium
modifications are rather small compared to the $\omega$-meson and
their production is suppressed due to the internal $s\bar{s}$ or
$c\bar{c}$ quark structure.

In heavy-ion collisions the baryon density varies dramatically
with time due to the formation and expansion of the
``fireball''. It was proposed in Refs.
~\cite{Sibirtsev1,Weidmann,Effenberger1,Effenberger2,Golubeva2,Sibirtsev2,schoen96,metag95,metag98}
to study the in-medium vector meson properties in
the interactions of pions, protons or photons with nuclei. Both
the $l^+l^-$ and $\pi^0\gamma$ invariant-mass spectra have been
discussed as possibilities investigating the vector-meson
properties in matter.

Only vector mesons decaying inside
nuclei can be used for an identification of the in-medium
properties. This imposes the kinematic
condition~\cite{Nikolaev} that the decay length of the vector
meson $L_V{\propto}E_V/m_V\Gamma_V$ should be less than the
nuclear radius. It implies that  the vector mesons should be
produced at small velocities or energies $E_V$ from heavy-nuclear
targets. Furthermore, because of the finite vector-meson
life time, $1/\Gamma_V$, only a fraction will decay inside the
nucleus and the final spectra will show the signal from decays
inside and outside the target nucleus as well as contributions
from the interference of the two amplitudes.

In case of  dilepton measurements in $\gamma A$ reactions
~\cite{Effenberger2} the in-medium signal from $\omega$-meson
decays is superimposed on a broad signal from the $\rho$-meson.
In addition, the dilepton measurements in
$\gamma A$ reactions are dominated by the Bethe-Heitler process,
which can only be suppressed by rather severe kinematic
cuts~\cite{Effenberger2}. Therefore, the $\pi^0\gamma$
measurements appear to be more promising.

The in-medium properties of vector mesons can be detected with
dilepton spectroscopy as mentioned above. As was proposed
in Refs.~\cite{Golubeva1,Sibirtsev1}, an alternative way to study
the in-medium modification of the $\omega$-meson is to
look at the $\omega{\to}\pi^0{+}\gamma$ decay mode. This mode
accounts for 8.5$\cdot$10$^{-2}$ of the  total $\omega$ decay width, while for
the $\rho$-meson it is only $6.8{\cdot}10^{-4}$.
The contribution of the $\rho$-meson to the $\pi^0\gamma$-mass distribution
is therefore negligible, which makes the $\pi^0\gamma$-decay mode an
exclusive probe to study the $\omega$-meson properties in matter.

It should be pointed out that the reconstruction of the
$\omega{\to}\pi^0{+}\gamma$ mode is hindered by the final-state
interactions of the $\pi^0$-meson in the nucleus, which might be
quite strong in case of large nuclei~\cite{Sibirtsev1}.
%%%%%%%%%%%%%%%%%%%%%%%%%%%%%%%%%%%%%%%%%%%%%%%%%%%%%%%%%%%%%%%%%
In addition, the pion from the $\omega$ decay may be off-shell,
too.
%%%%%%%%%%%%%%%%%%%%%%%%%%%%%%%%%%%%%%%%%%%%%%%%%%%%%%%%%%%%%%%%%%
 Furthermore, the finite experimental efficiency causes a
mis-identification of four $\gamma$ events from the channels
$\pi^0\pi^0$, $\eta\pi^0$ and related reactions with more than
three photons in the final channel. Thus the $\pi^0 \gamma
\rightarrow 3 \gamma$ mode requires a full understanding of the
contributions from these reactions.

We investigate the possibility to deduce the in-medium
modification of the $\omega$-meson from the $\pi^0\gamma$ invariant-mass
distribution for $\gamma{+}A$ reactions and evaluate the optimal experimental
conditions for the photon energy and the size of the target
nucleus. We will apply kinematic cuts on 3 photon
events in order to identify the distortion of the in-medium signal
due to the $\pi^0$-meson final state interaction. Our study is
designed for the TAPS~\cite{Gabler} and Crystal Barrel~\cite{aker92}
experiment at ELSA. The detector response is known and
the background contribution due to the final-state mis-identifications,
caused by the detector efficiency and the corrections due to the
experimental resolution in the reconstruction of the $\pi^0\gamma$
spectrum, are calculated.

\section{Ingredients of the  model}
\label{sec:1}

\begin{sloppypar}
The calculations were performed within the transport model previously
applied to $\rho$, $K^+$, $K^-$ and $\omega$ meson
production in pion-nucleus and proton-nucleus
collisions~\cite{Sibirtsev1,Sibirtsev2,Sibirtsev3}.
In the following, we describe the new ingredients of the model relevant to
photoproduction of $\omega$-mesons and the calculation of the
background $\gamma{+}N{\to}\pi^0{+}\pi^0{+}N$ reaction.
\end{sloppypar}

\begin{figure}[h]
\vspace{-7mm}\hspace{-7mm}
%\vspace{10.cm}
\resizebox{0.5\textwidth}{!}{\includegraphics{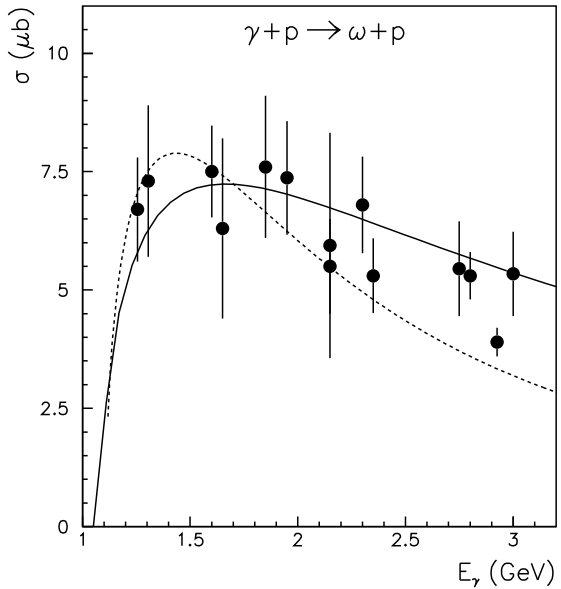}}
\vspace{-2mm}
\caption{The $\omega$-meson photoproduction cross section on the
proton as a function of the photon energy. The experimental data
are taken from Ref.~\cite{Landolt}. The solid line shows the
parameterization~(\ref{fitt1}), while the dashed line corresponds
to a calculation within the $\pi$- and $\eta$-meson exchange
model~\cite{Friman3,Titov}.} \label{gres2-a} \vspace{-1mm}
\end{figure}

In Fig.~\ref{gres2-a}, the $\gamma{+}p{\to}\omega{+}p$
cross section is shown as a function of the photon energy $E_\gamma$ from
Ref.~\cite{Landolt}. The dashed line gives the result for the
$\omega$ cross section calculated within the $\pi$ and
$\eta$-meson exchange model where the parameters have been adopted
from Refs.~\cite{Friman3,Titov}. Our calculations
are close to the predictions of Refs.~\cite{Friman3,Titov}. The
discrepancy between the data and the calculations at
$E_\gamma{\ge}2.5$~GeV might be due to additional contributions
from pomeron exchange~\cite{Titov}, which plays a dominant role at
high energies~\cite{Donnachie}.

For an application in our transport calculations we parameterize
the data on the $\gamma{+}p\to\omega{+}p$ cross section as
\begin{equation}
\sigma{=}\frac{[(s{-}m_N^2{-}m_\omega^2)^2{-}4m_N^2m_\omega^2]^{1/2}}
{s \, (s-m_N^2)} \frac{A}{(\sqrt{s}{-}M)^2{+}\Gamma^2/4},
\label{fitt1}
\end{equation}
where $s$, $m_N$ and $m_\omega$ are the squared invariant
collision energy, the nucleon and $\omega$-meson masses, respectively.
The parameters $M{=}1.7$~GeV, $\Gamma$=6~GeV and
$A$=396~$\mu$b$\cdot$GeV$^{-4}$ have been fitted to the
data~\cite{Landolt}.

The solid line in Fig.~\ref{gres2-a} shows the
parametrization~(\ref{fitt1}) that reproduces the available
data~\cite{Landolt} reasonable well. We mention that the
parameterization~(\ref{fitt1}) is also in agreement with recent
SAPHIR data~\cite{Klein1,Klein2}. The cross section on the neutron
was taken to be the same as on a proton.

In the transport code the $\omega$-mesons are produced in
accordance with their spectral function. In addition to the vacuum
spectral function, an in-medium modification was introduced by the
real and imaginary part of an effective $\omega$-meson
potential~\cite{Sibirtsev1}
\begin{eqnarray}
\Re U_\omega (\rho) = m_\omega \ \beta \ \frac{\rho}{\rho_0}, \nonumber \\
\Im U_\omega(\rho) =\Gamma_{coll} \  \frac{\rho}{\rho_0},
\label{dens}
\end{eqnarray}
\begin{sloppypar}
\noindent where $m_\omega$ is the bare $\omega$-meson mass, $\rho_0$=0.16
fm$^{-3}$ and $\rho$ is the local nuclear density. The parameter
$\beta{=}$--0.16 was adopted from the model predictions in
Refs.~\cite{Tsushima1,Klingl1,Friman2,Saito2,Saito3,Klingl2} and
the collisional width $\Gamma_{coll}$=50~MeV was taken from
Refs.~\cite{Klingl1,Lykasov}. We calculate the production,
the propagation and decay of
the $\omega$-meson inside the potential~(\ref{dens}). This
prescription consistently guarantees that the $\omega$-mesons
decaying outside the nuclei regain the $\omega$-spectral
function in vacuum.
\end{sloppypar}

\begin{sloppypar}
The linear-density approximation in Eq.~(\ref{dens}) is in
reasonable agreement\footnote{For instance, the comparison between
the scaling~\cite{Brown,Hatsuda1} and the
calculations~\cite{Saito1,Tsushima2} for a carbon target is shown
in Fig.~3 of Ref.~\cite{Sibirtsev3}.} with the density dependence
of the real part of the vector-meson potential calculated in
Refs.~\cite{Saito1,Tsushima2} for finite nuclei.
\end{sloppypar}

Furthermore, we account for the elastic and inelastic
$\omega$-meson interactions in nuclei employing the cross sections
given in Ref.~\cite{Lykasov}. We calculate the distortion of the
$\omega{\to}\pi^0\gamma$ signal due to the re-scattering of the
$\pi^0$-meson by adopting the total and elastic $\pi{+}N$
differential cross sections from the Karlsruhe-Helsinki
partial-wave analysis~\cite{Hohler,Koch}.
%%%%%%%%%%%%%%%%%%%%%%%%%%%%%%%%%%%%%%%%%%%%%%%%%%%%%%%%%%%%%%%
In the medium, however, many-body phenomena like two-nucleon
absorption of pions may occur additionally, which is very
pronounced for low momentum pions \cite{ericson}, but becomes
questionable for pion momenta above 350 MeV/c since the pion
wavelength must be somewhat larger than the average distance
between nucleons. We have investigated the role of such many-body
effects by calculating $\pi^+$ absorption on nuclei for pion
momenta above 300 MeV/c employing the cross sections
from~\cite{Hohler,Koch}. As a genuine result we find that for
$^{65}Fe$ and $^{209}Bi$ nuclei we overestimate the pion
absorption by 10-20 \%, respectively, in comparison to the data
from Ref. \cite{piabs}. This result is fully in line with the
earlier transport studies by Engel et al. in Ref. \cite{Engel}.
In addition, we have analyzed the two-nucleon absorption in the
$\omega$-production channel discussed in this paper. For the
energies presented here, the effect of an additional 
two-nucleon absorption is estimated to be around 5\%.
We thus discard additional many-body phenomena 
in order not to overestimate the $\pi^0$ reabsorption.
%%%%%%%%%%%%%%%%%%%%%%%%%%%%%%%%%%%%%%%%%%%%%%%%%%%%%%%%%%%%%%%%%%

\begin{figure}[h]
\vspace{-7mm}\hspace{-7mm}
%\vspace{10cm}
\resizebox{0.5\textwidth}{!}{\includegraphics{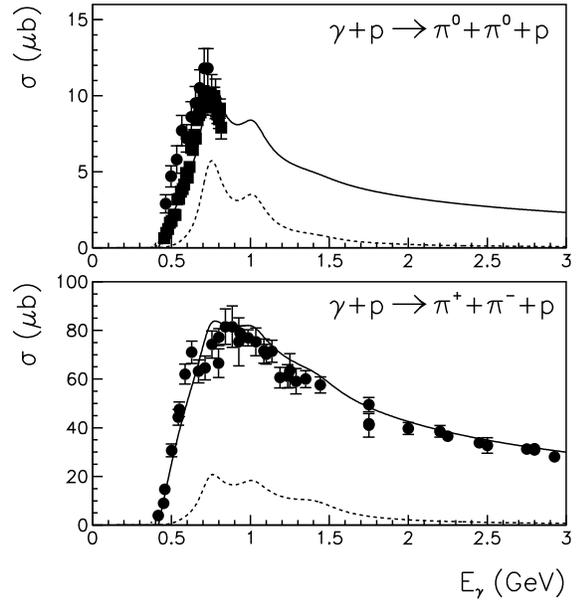}}
\vspace{-2mm}
\caption{The $\gamma{+}p{\to}\pi^0{+}\pi^0{+}p$ and
$\gamma{+}p{\to}\pi^+{+}\pi^-{+}p$ cross sections as a function
of the photon energy. The experimental data are taken from
Refs.~\cite{wolf,Landolt}. The solid lines show our parameterization,
while the dashed lines indicate the contribution from the baryonic
resonances.} \label{resbro7a} \vspace{-1mm}
\end{figure}

To calculate the background from two neutral pions we implement
the elementary photoproduction cross section shown in
Fig.~\ref{resbro7a}. While data~\cite{Landolt} for the
$\gamma{+}p{\to}\pi^+{+}\pi^-{+}p$  cross section are available for
a large range of photon energies, data for the
$\gamma{+}p{\to}\pi^0{+}\pi^0{+}p$ reaction are only available
up to $E_\gamma{\simeq}$0.8 GeV~\cite{wolf}. We employ a resonance model to
construct a parameterization over the desirable range of photon
energies. As shown by the dashed lines in Fig.~\ref{resbro7a},
the contribution from the baryonic resonances coupled to two pions
can be calculated for both reactions by taking into account the
resonance properties from the analysis in
Ref.~\cite{Manley}. We assume the same energy dependence of the
non-resonant contribution for the
$\gamma{+}p{\to}\pi^0{+}\pi^0{+}p$ and the $\gamma{+}p{\to}\pi^+{+}\pi^-{+}p$
reaction, however, scale the strength in line with the low-energy
data~\cite{wolf}. Our estimate for the $\gamma{+}p{\to}\pi^0{+}\pi^0{+}p$
cross section is in reasonable agreement with the microscopic
calculations of Ref.~\cite{Gomez}.

Experimental data on the $\gamma{+}N{\to}\eta{+}\pi^0{+}N$
reaction, that might contribute to the background, are not
available. We thus discard this channel in our present
calculations. It is known experimentally~\cite{Landolt}
that the total $\gamma{+}N$ cross section is
almost saturated by the sum of the $\gamma{+}N{\to}\pi{+}N$,
$\gamma{+}N{\to}2\pi{+}N$ and $\gamma{+}N{\to}\eta{+}N$ partial
cross sections below a photon energy of 1.3~GeV. At higher photon
energies the dominant contributions to the total $\gamma{+}N$
cross section stem from multi-pion production~\cite{Landolt}.

%%%%%%%%%%%%%%%%%%%%%%%%%%%%%%%%%%%%%%%%%%%%%%%%%%%%%%%%%%%%%%%%%%%
Since not only the $\omega$-meson, but also the decay pion changes
its spectral function in the medium, the in-medium $\omega$ Dalitz
decay has to be discussed explicitly. Now let the in-medium
$\omega$ mass be $M$ and the decay pion have a mass $m^*_\pi$
that is selected by Monte Carlo according to its in-medium width
$\Gamma_\pi^*$ and mass shift $\delta m_\pi$. Energy and momentum
conservation - in the rest frame of the $\omega$ meson - than
implies
\begin{equation}
M^2 = (E_\pi^*+E_\gamma)^2 = (\sqrt{p_{\pi}^{*2} + m_{\pi}^{*2}} +
E_\gamma)^2  \label{e1},
\end{equation}
with $E_\gamma = |p_\pi^*|$ denoting the photon energy which in
magnitude equals the momentum of the decay pion. Whereas the
photon propagates to the vacuum without distortion, the in-medium
pion changes its momentum and spectral function during the
propagation to the vacuum according to quantum off-shell
propagation \cite{ca1,ca2}. In the particular case, where the pion
self energy $\Sigma_\pi$ has no explicit time dependence
($\partial_t \Sigma_\pi = 0$) and is only a function of momentum
and density, which holds well for the case of $p + A$ reactions,
the energy of the pion $E_\pi^* = E_\pi$ is a constant of motion
(cf. Eq. (20) in Ref. \cite{ca2}). Thus the off-shell mass and
momentum balance out during the propagation as shown graphically
in Ref. \cite{ca2} in Figs. 1 and 2 for a related problem. The
magnitude of the pion momentum in vacuum - if not scattered
explicitly or being absorbed - then is  given by
\begin{equation}
p_\pi^v = \sqrt{p_\pi^{*2} + m_\pi^{*2} - m_0^2}, \label{e2}
\end{equation}
where $m_0$ denotes the vacuum pion mass and $p_\pi^*$ the pion
momentum in the $\omega$ decay in-medium. The invariant mass
(squared) of the pion and photon in the vacuum then is given by
\begin{equation}
M_r^2 = (E_\pi + E_\gamma)^2 - (p_\pi^v - E_\gamma)^2 = M^2 -
(p_\pi^v - E_\gamma)^2 . \label{e3}
\end{equation}
The pion propagation thus leads to a slight downward shift of the
invariant mass in vacuum relative to its original in-medium value.
We note, that the above considerations apply well for heavy nuclei
where the corrections in energy due to the recoil momentum of the
nucleus ($|p_\pi^v - E_\gamma|$) can be discarded. 

\begin{figure}[h]
\vspace{-2mm}\hspace{2mm}
%\vspace{10cm}
\resizebox{0.45\textwidth}{!}{\includegraphics{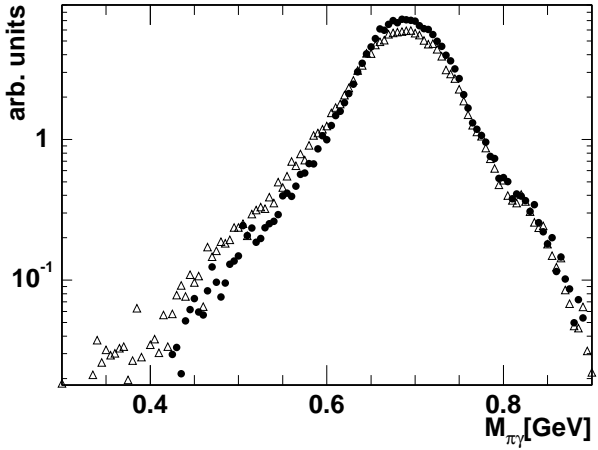}}
\vspace{-2mm}
\caption{A BUU Monte-Carlo prediction of the in-medium $\pi^0\gamma$ 
mass distribution for the reaction $\gamma+Ca\rightarrow \pi^0\gamma+X$ 
at an incident photon energy of 1.2~GeV. The effect of the mean-field 
interaction of the $\pi^0$ with the nucleus is studied by comparing 
the prediction of an unmodified $\pi^0$ meson (filled circles) and 
the prediction for which the $\pi^0$ mass is modified according
to a Breit-Wigner distribution with a width of 150~MeV (open triangles).}
\label{meanField-fig} \vspace{-1mm}
\end{figure}

For an actual quantification of this effect one has to specify the
pion spectral function e.g. at nuclear matter density $\rho_0$.
The pions, that originate from the $\omega$ decay in $\gamma+A$
reactions, have typical momenta $\geq 300$ MeV such that their
interaction cross section with nucleons is in the order of 30-35
mb, while their velocity $\beta_\pi$ relative to the target is
close to 1. Consequently the collisional broadening can be
estimated as
\begin{equation}
\delta \Gamma_\pi \approx \beta_\pi \sigma _{\pi N} \rho_0 \approx
85 - 120 \ MeV, \label{e4} \end{equation} which is still quite
substantial. For a test case we have assumed the $\omega$
in-medium spectral function to be represented by a Breit-Wigner
of width $\Gamma_{tot}$ = 50 MeV and pole mass of 0.65 GeV. Then
for each invariant mass $M$ the decay of the $\omega$ to a photon
and an in-medium pion can be evaluated and the pion momentum
$p_\pi^v$ according to (\ref{e2}). The results of the simulations
for the pion off-shell effects on the reconstructed invariant
mass distribution in (\ref{e3}) show that for a pion collisional
broadening of 100 -- 200 MeV the shift in invariant mass is
practically invisible at the pole mass and only becomes more
pronounced in the low mass tails of the invariant mass
distribution. In Fig.~\ref{meanField-fig} the $\pi^0$ off-shell effect 
is displayed quantitatively for the reaction 
$\gamma + Ca\rightarrow \pi^0\gamma + X$ 
at an incident photon energy of 1.2~GeV. Shown is the $\pi^0\gamma$-mass 
distribution for those $\omega$ mesons which decay inside the nucleus. 
Indicated are the distributions assuming freely propagating on-shell 
$\pi^0$ mesons (filled circles) and assuming a modification of the 
$\pi^0$ spectral function by a Breit-Wigner distribution with a width 
of $\Gamma$=150~MeV (open triangles). 
As seen from Fig.~\ref{meanField-fig}, the mean-field propagation of 
the $\pi^0$ meson in the nucleus 
can be safely neglected especially in view of presently 
achievable experimental mass resolutions 
(see Section~\ref{experimental-resolution}). 
Furthermore, the spectrum is strongly
distorted by $\pi^0$ rescattering only in the low mass tails 
(see Section~\ref{sec:pion-rescat}).
Thus we continue our studies with on-shell calculations 
for the pions without substantial loss in accuracy.
%%%%%%%%%%%%%%%%%%%%%%%%%%%%%%%%%%%%%%%%%%%%%%%%%%%%%%%%%%%%%%%%%%%

\section{Model predictions}
\label{predictions}

The coupled-channel transport model described in the previous section
has been adopted in a Monte-Carlo simulation to examine and optimize
the feasibility of measuring the $\omega$ mass at nuclear densities
in the process $\gamma+A\rightarrow \pi^0\gamma +X$.
In particular, the effect of the final-state
$\pi^0$ rescattering to the $\omega$-mass determination is studied.
Furthermore,
the dependence on the incident photon energy and the atomic mass of
the nucleus is addressed in this section.
In sections~\ref{experimental-resolution} and
\ref{experimental-background} the effect of experimental resolution
and background processes are discussed.

\subsection{Incident photon energy}

\begin{figure}
\resizebox{0.5\textwidth}{!}{\includegraphics{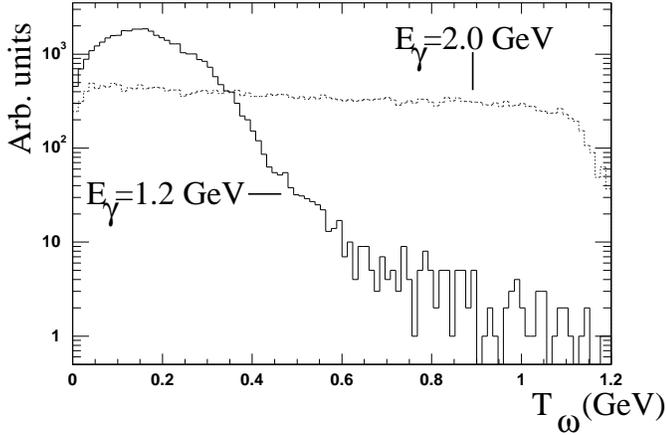}}
\caption{The kinetic-energy distributions of the $\omega$ meson in the reaction
$\gamma+Ca\rightarrow \omega+X$ for incident photon energies of
$E_{\gamma}$=1.2~GeV (solid histogram) and $E_{\gamma}$=2.0~GeV
(dotted histogram).}
\label{fig:photon_energies}
\end{figure}

\begin{sloppypar}
The fraction of $\omega$ mesons decaying inside the nucleus can be
optimized by minimizing the decay length
$L_\omega{\propto}E_\omega/m_\omega\Gamma_\omega$.
It is therefore preferred that the kinetic energy of the $\omega$ meson is small.
This can be achieved with an incident photon energy close to
the $\omega$-production threshold as illustrated in
Fig.~\ref{fig:photon_energies}. Here, the kinetic-energy
distributions of the $\omega$ produced in the reaction
$\gamma+Ca\rightarrow \omega +X$ for an incident photon energy
of $E_{\gamma}$=1.2~GeV (solid histogram) and an incident photon energy
of $E_{\gamma}$=2.0~GeV (dotted histogram) are compared.
Clearly, the $\omega$ meson has the smallest average energy for
$E_{\gamma}$=1.2~GeV. Furthermore, at $E_{\gamma}$=1.2~GeV
the $\omega$-production cross section is still comparable to the cross sections at
larger incident photon energies (see Fig.~\ref{gres2-a}).
Thus in the simulations described below an incident photon
energy of $E_{\gamma}$=1.2~GeV is used.
\end{sloppypar}

\subsection{$\pi^0$ rescattering}
\label{sec:pion-rescat}

\begin{figure}
\resizebox{0.5\textwidth}{!}{\includegraphics{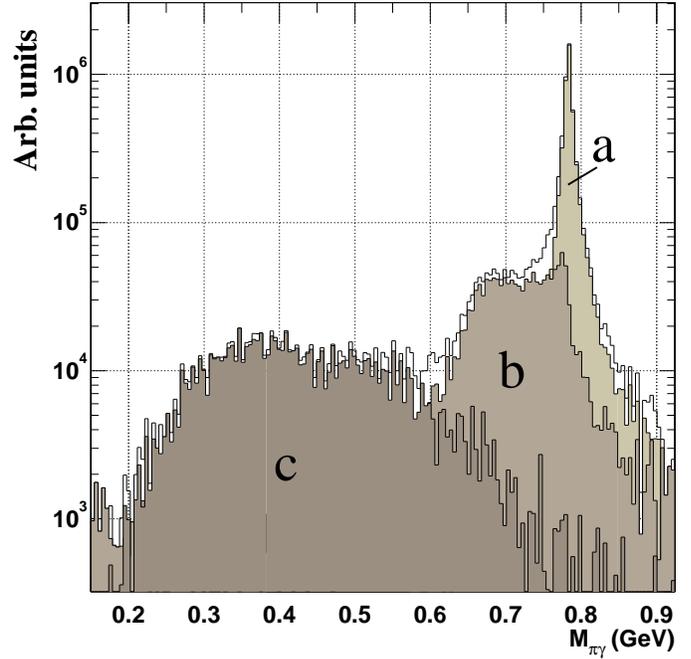}}
\caption{The $\pi^0\gamma$-mass distribution obtained from a Monte-Carlo
simulation of the process $\gamma+Nb\rightarrow \pi^0\gamma+X$ at
$E_{\gamma}$=1.2~GeV. The spectrum is decomposed into different
contributions corresponding to the fraction of $\omega$ mesons decaying
outside ($\rho<0.05$~fm$^{-3}$) the nucleus ({\bf a}),
the fraction of $\omega$ mesons decaying inside ($\rho>0.05$~fm$^{-3}$)
for which the $\pi^0$ does not rescatter ({\bf b}), and the fraction of
$\omega$ mesons decaying inside the nucleus for which $\pi^0$
rescatters ({\bf c}).}
\label{fig:nb_decompose}
\end{figure}

\begin{sloppypar}
One of the critical arguments against studying the $\omega$-mass in-medium
by the decay $\omega\rightarrow \pi^0\gamma$ is the strong final-state
interaction of the $\pi^0$.
The transport model allows to study this effect quantitatively.
In Fig.~\ref{fig:nb_decompose} this is illustrated for the reaction
$\gamma+Nb\rightarrow\pi^0\gamma+X$ at $E_{\gamma}$=1.2~GeV. The figure shows
the $\pi^0\gamma$-mass distribution decomposed in the fraction of $\omega$
mesons decaying outside the nucleus ({\bf a}), the fraction decaying inside
the nucleus ({\bf b}) (for which the $\pi^0$ does not rescatter) and the fraction
decaying inside the nucleus in case of $\pi^0$ rescattering ({\bf c}).
Note, that a fraction of approximately 36\% of the
$\omega$ mesons decays inside the nucleus.
By definition, a particle decays ``inside'' the nucleus if the density
$\rho>0.05$~fm$^{-3}$.
A shift (of $\approx$20\%) and the effect of collisional
broadening (of $\approx$50~MeV) of the $\omega$-mass distribution can be
observed as parametrized according to Eqs.~(\ref{dens}).
For $\approx$40\% of all the $\omega$ mesons decaying inside the nucleus,
the final-state $\pi^0$ does rescatter. However, this background appears
dominantly at relatively small invariant masses $M_{\pi^0\gamma}$
and is spread out over a large mass range. The scattering of the $\pi^0$ with
the nucleons alter its direction and
kinetic energy such that the initial $\omega$-mass information is
essentially lost. Hence, the $\pi^0$-rescattering contribution is
significantly smaller in the range 0.6$<M_{\pi^0\gamma}<$0.8~GeV
than the contribution of $\omega$ mesons decaying inside without
$\pi^0$ rescattering.
\end{sloppypar}

\begin{figure}
\resizebox{0.5\textwidth}{!}{\includegraphics{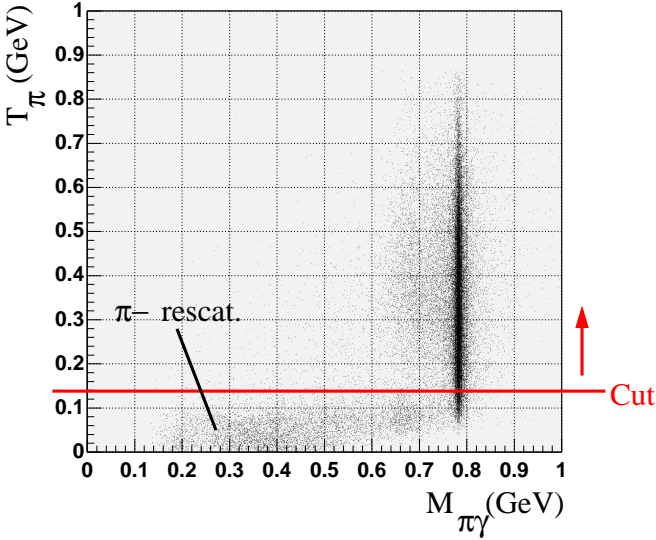}}
\caption{The $\pi^0\gamma$-mass distribution versus the reconstructed
kinetic energy of the final-state $\pi^0$.}
\label{fig:e_pion_vs_m_omega}
\end{figure}

\begin{figure}
\resizebox{0.45\textwidth}{!}{\includegraphics{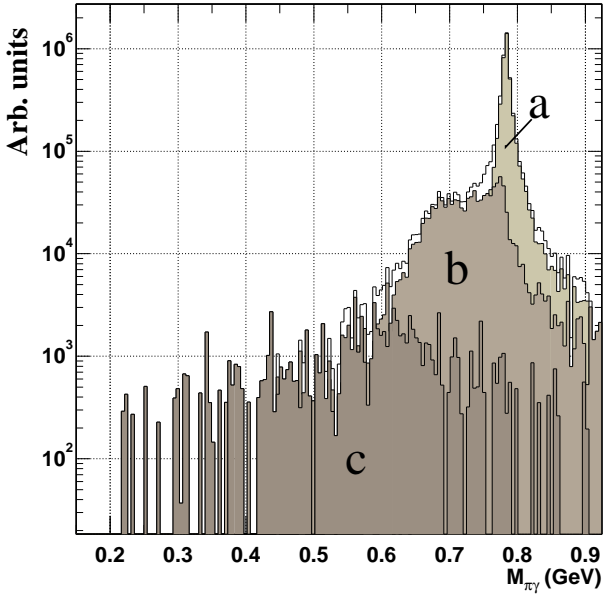}}
\caption{The same as Fig.~\ref{fig:nb_decompose} with the additional condition
of $T_{\pi^0}>150$~MeV.}
\label{fig:nb_decompose_cut}
\end{figure}

\begin{sloppypar}
The energy loss of the rescattered $\pi^0$ mesons can be
exploited experimentally to reduce the background from rescattering even further. This
is illustrated in Fig.~\ref{fig:e_pion_vs_m_omega}, showing the
$\pi^0\gamma$-mass distribution versus the reconstructed kinetic energy
of the $\pi^0$. Clearly, the contribution of $\omega$ mesons decaying
inside the nucleus - with $\pi^0$ rescattering - is correlated with
low-energy $\pi^0$. Without a significant loss of the actual signal
($\omega$ mesons decaying inside the medium without rescattering of
the $\pi^0$), the dominant part of the $\pi^0$-rescattering background
can be eliminated by selecting $\pi^0$ mesons with a kinetic energy
$T_{\pi^0}>150$~MeV. The resulting $\pi^0\gamma$-mass distribution
is shown in Fig.~\ref{fig:nb_decompose_cut}. An improvement
compared to Fig.~\ref{fig:nb_decompose} can be observed. The contribution
of the $\pi^0$-rescattering background is now reduced to 1\% within
the mass range of $0.6<M_{\pi^0\gamma}<0.8$~GeV.
%%%%%%%%%%%%%%%%%%%%%%%%%%%%%%%%%%%%%%%%%%%%%%%%%%%%%%%%%%%%%%%%
Besides exploiting the energy loss of the rescattered $\pi^0$, 
angular correlations between the 
incident photon, the final-state photon and the $\pi^0$ meson can be used
to enhance the in-medium signal relative to $\pi^0$-rescattering background. 
An analysis has shown that a similar result as shown in 
Fig.~\ref{fig:nb_decompose_cut} can be obtained by alternatively gating
on the opening angle between the final-state $\pi^0$ and $\gamma$ and
cutting on the out-of-plane angle of the $\pi^0$ relative to the incident
and final-state $\gamma$-$\gamma$ plane.
%%%%%%%%%%%%%%%%%%%%%%%%%%%%%%%%%%%%%%%%%%%%%%%%%%%%%%%%%%%%%%%%
\end{sloppypar}

\subsection{Atomic-mass dependence}

\begin{figure}
\resizebox{0.45\textwidth}{!}{\includegraphics{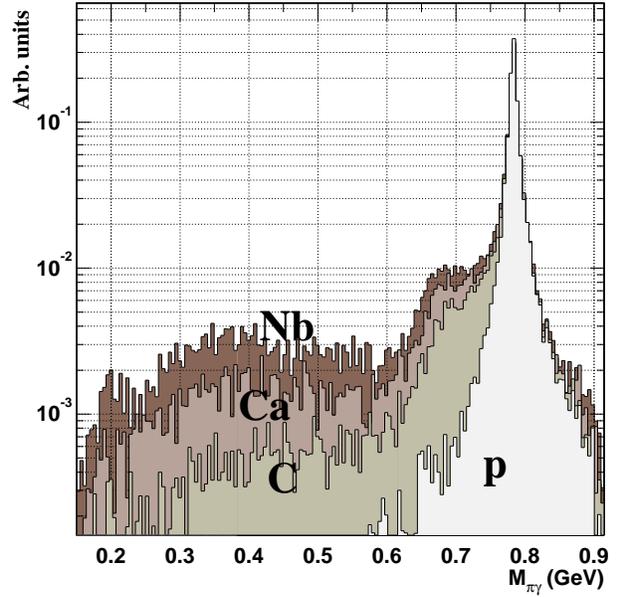}}
\caption{The $\pi^0\gamma$-mass distribution as predicted by the
coupled-channel transport model for $\gamma+A\rightarrow \pi^0\gamma+X$
at $E_{\gamma}$=1.2~GeV and $A$=$p$, $C$, $Ca$ and $Nb$. The spectra
are normalized at the $\omega$ peak $M_{\pi^0\gamma}$=0.78~GeV.}
\label{fig:diftargets_nores}
\end{figure}

The dependence of the $\omega$-mass in-medium on the atomic mass
($A$) has been studied by performing Monte-Carlo simulations for
different nuclei. The result is shown in
Fig.~\ref{fig:diftargets_nores} for a proton, carbon, calcium and
niobium target, varying $A$ between 1 and 92. All spectra are normalized
at the pole mass of the $\omega$ ($M_{\pi^0\gamma}$=0.78~GeV).
Obviously, the contribution of the $\omega$ mesons decaying inside
the nucleus (with and without $\pi^0$ rescattering) increases as
function of $A$ simply due to an increase in the effective radius of
the nucleus. One might argue that a larger nucleus, like $Pb$,
is more suited. However,
for heavy targets one has to take into account
the decrease in radiation length with $Z^2$, therefore
increasing the fraction of incident photons converting into a lepton pair.
In Table~\ref{tab:diftargets} we have summarized a
quantitative analysis of the different contributions for different targets,
e.g. the fraction of $\omega$ mesons decaying outside the nucleus,
inside the nucleus without $\pi^0$ rescattering, and inside
the nucleus with $\pi^0$ rescattering. In addition, the effect of a
cut on $T_{\pi^0}>150$~MeV is included in the table (values in brackets).

\begin{table}
\caption{The fraction of $\omega$ mesons decaying outside the nucleus, inside
the nucleus (i.e. at $\rho>0.3\,\rho_0$) with and without $\pi^0$ rescattering for carbon, calcium and
niobium. The effect of an additional cut on $T_{\pi^0}>$150~MeV is shown
as the values within brackets. All values are given in percentage of the
total $\omega$-production rate.}
\label{tab:diftargets}
\begin{tabular}{llccc}
\hline\noalign{\smallskip}
Nucleus & Decay & Decay inside & Decay inside\\
& outside & without $\pi^0$ & with $\pi^0$\\
& & rescattering & rescattering \\
\noalign{\smallskip}\hline\noalign{\smallskip}
$C$ ($A$=12) & 83.7 (86.1) & 13.4 (13.3) & 2.8 (0.62) \\
$Ca$ ($A$=40) & 73.8 (80.4) & 18.4 (18.4) & 7.8 (1.2) \\
$Nb$ ($A$=92) & 65.0 (75.7) & 21.5 (22.0) & 13.5 (2.3) \\
\noalign{\smallskip}\hline
\end{tabular}
\end{table}

\section{Experimental resolution}
\label{experimental-resolution}

\begin{figure}
\resizebox{0.45\textwidth}{!}{\includegraphics{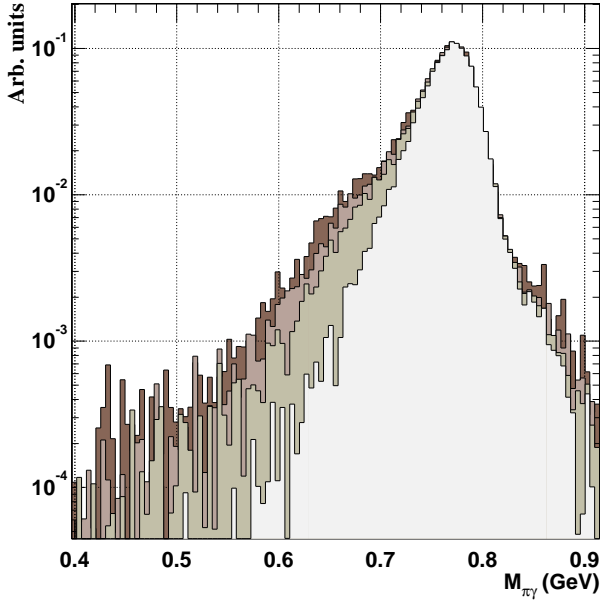}}
\caption{The $\pi^0\gamma$-mass distribution corresponding to
Fig.~\ref{fig:diftargets_nores} for $p$, $C$, $Ca$, $Nb$ including the effect of a realistic
detector resolution. Furthermore, a cut on the kinetic energy of
the pion of $T_{\pi^0}>$150~MeV is applied.}
\label{fig:diftargets_res}
\end{figure}

A realistic simulation has to take the effect of a finite
detector resolution on the $\pi^0\gamma$-mass determination into account. Therefore,
we have included in the simulation a photon position resolution of
$\sigma$=1.2$^\circ$ and a photon energy response as
measured with the photon spectrometer TAPS consisting of BaF$_2$
crystals~\cite{Gabler}. A similar resolution is expected for the
Crystal Barrel detector ($CsI$)~\cite{aker92}, therefore providing a realistic simulation
for future experiments at ELSA. The effect on the invariant-mass
determination for different nuclei is shown in Fig.~\ref{fig:diftargets_res}.
Although an obvious decrease in the mass resolution can be
observed by a comparison with Fig.~\ref{fig:diftargets_nores}, a
shoulder due to $\omega$ mesons decaying inside the nucleus
is still visible and increases with the size of the nucleus.
The figure demonstrates the necessity of measuring
the $\pi^0\gamma$ mass for a set of different targets to
avoid experimental ambiguities.

\section{Channel misinterpretation}
\label{experimental-background}

\begin{figure}
\resizebox{0.45\textwidth}{!}{\includegraphics{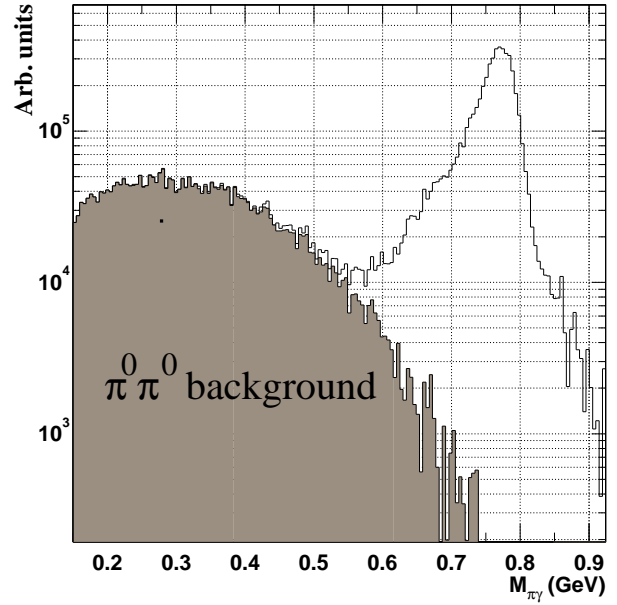}}
\caption{The $\pi^0\gamma$-mass distribution as expected for $\gamma+Nb$ at
an incident photon energy of 1.2~GeV. The contribution from the background
channel $\gamma+Nb\rightarrow \pi^0\pi^0+X$ is indicated as grey area.
The probability of not detecting a photon is assumed to be 5\%. Furthermore,
the detector response is included and a cut on $T_{\pi^0}>150$~MeV is applied.}
\label{fig:two-pion-nocut}
\end{figure}

So far, the effect of experimental background has not been considered
in the simulations. We expect the main source of background to
come from a misidentification of the $\gamma+A\rightarrow \pi^0\gamma +X$
reaction due to not detecting one of the photons of the process
$\gamma+A\rightarrow \pi^0\pi^0+X$. Therefore, the latter reaction
has been implemented in the calculation as described in
section~\ref{sec:1} and simulated in the same framework as the
$\gamma+A\rightarrow\pi^0\gamma+X$ process. The result for
$\gamma+Nb$ at an incident photon
energy of 1.2~GeV is shown in Fig.~\ref{fig:two-pion-nocut}. Here, we assumed
that the probability of not detecting a photon due to detector thresholds,
inefficiencies and acceptances is 5\%. The contribution of the
$\pi^0\pi^0$ background as indicated by the shaded part of the
spectrum is $\approx$36\% of the total yield. In the mass range
of interest ($0.6<M_{\gamma\pi^0}<0.8$~GeV) the contribution is only
1\%.

\begin{figure}
\resizebox{0.5\textwidth}{!}{\includegraphics{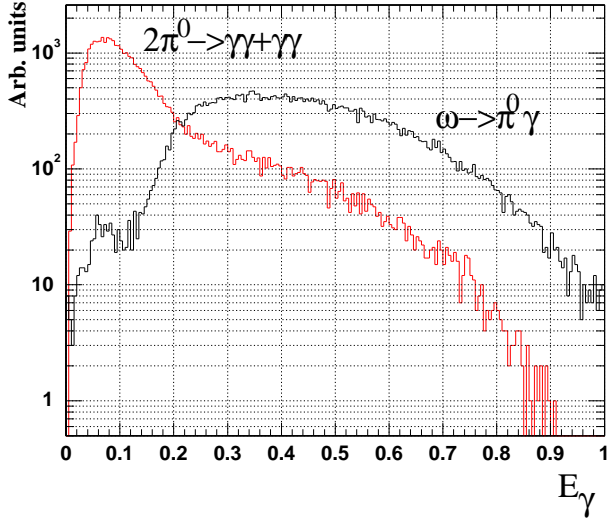}}
\caption{The energy distributions of the photons (that do not correspond to one of
the decay products of the pion) from the $\gamma+Nb\rightarrow \pi^0\gamma +X$
channel is compared with the energy distribution of photons from the
$\gamma+Nb\rightarrow \pi^0\pi^0+X$ channel.}
\label{fig:two-pion-cut}
\end{figure}

\begin{figure}
\resizebox{0.45\textwidth}{!}{\includegraphics{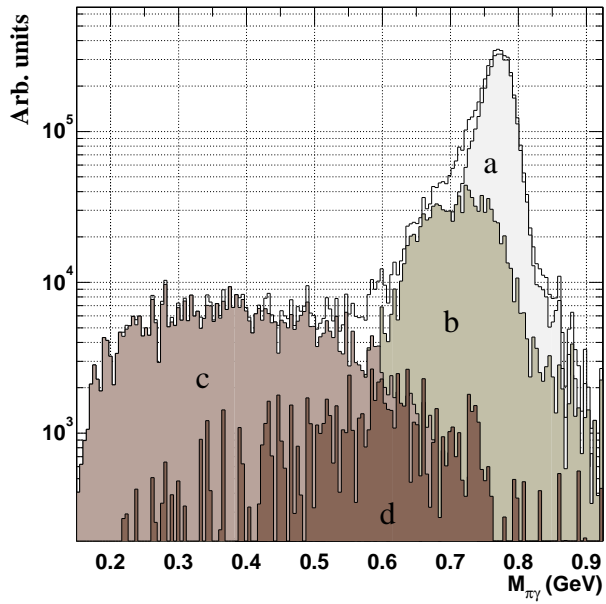}}
\caption{The $\pi^0\gamma$-mass distribution for $\gamma+Nb$ at an
incident photon energy of 1.2~GeV. The fraction of $\omega$ mesons
decaying outside the nucleus ({\bf a}), inside the nucleus without $\pi^0$
rescattering ({\bf b}), inside the nucleus with $\pi^0$ rescattering
({\bf d}) and the background from the 2$\pi^0$ ({\bf c}) are indicated.
The experimental resolution
is taken into account. Furthermore, kinematical cuts to reduce the
background from $\pi^0$ rescattering and from the 2$\pi^0$ channel
have been applied (see text).}
\label{fig:final}
\end{figure}

The fraction of the $\pi^0\pi^0$ background to the total mass distribution
can be reduced by simple kinematic cuts as proposed in
Fig.~\ref{fig:two-pion-cut}. In this figure,
the energy distribution of one of the photons of the
$\gamma+Nb\rightarrow\pi^0\pi^0+X$ reaction (grey line) is compared to
the energy  distribution of the photon of the
$\gamma+Nb\rightarrow\pi^0\gamma+X$ reaction (solid line).
Clearly, requiring the photon energy to be larger than 200~MeV reduces
the $\pi^0\pi^0$ background significantly without a large loss of the
actual signal. The effect of an additional cut of $E_{\gamma}>200$~MeV
on the $M_{\pi^0\gamma}$ distribution is shown in Fig.~\ref{fig:final} for
$\gamma+Nb$ at an incident photon energy of 1.2~GeV.
In addition, the contribution of $\omega$ mesons decaying outside
the nucleus and the contribution decaying inside the nucleus,
with and without $\pi^0$ rescattering, are depicted. Note that the
background from $\pi^0$ rescattering and from the $\pi^0\pi^0$ channel
within the complete mass range between $M_{\pi^0\gamma}$=0.6 and 0.8~GeV
is significantly smaller than the signal of interest, i.e. the fraction
of $\omega$ mesons decaying inside without a rescattering of the $\pi^0$.

\section{Conclusions}

In this study the feasibility of measuring the $\omega$ spectral function at normal
nuclear densities is discussed by exploiting the process $\gamma+A\rightarrow\pi^0\gamma+X$.
Experimentally, this process will be studied with the Crystal Barrel
detector~\cite{aker92} at ELSA in coincidence with the
photon spectrometer TAPS~\cite{Gabler}.
Such experiments are complementary to plans at GSI and CEBAF to
measure the channels $\pi^-+A\rightarrow e^+e^-+X$ and
$\gamma+A\rightarrow e^+e^-+X$ with the HADES~\cite{Fri99} and
CLAS~\cite{clas} detector, respectively.
The processes are sensitive to the spectral functions of the $\rho$-meson,
which has a large decay branch into $e^+e^-$ pairs. Furthermore, photon-induced
dilepton studies have to take into account the contribution from the dominant
Bethe-Heitler process.

%Final-state interaction pion

\begin{sloppypar}
The main drawback of measuring the $\omega$ mass in-medium via its decay
$\omega\rightarrow\pi^0\gamma$ is the use of a hadronic probe ($\pi^0$) which
has a sizeable probability to interact with the nucleons. The influence of this
background on the mass determination has been studied using a coupled-channel
transport calculation adopted in a Monte-Carlo simulation. Although the final-state
interactions of the $\pi^0$ meson are found to be large, its contribution
at the invariant mass region of interest ($0.6<M_{\pi^0\gamma}<0.8$~MeV) is
small. A kinematic cut on the kinetic energy of the $\pi^0$ allows for a further
reduction of the $\pi^0$-rescattering background, leading to an almost negligible
contribution to the $\pi^0\gamma$ invariant-mass distribution.
\end{sloppypar}

%Experimental resolution and background

The photon spectrometers, TAPS and Crystal Barrel, have a finite mass
resolution which is larger than the decay width of a free-propagating $\omega$ meson.
This hinders the observation of an in-medium mass shift of the $\omega$ meson.
Simulations including a realistic $\pi^0\gamma$-mass resolution show that within the
assumption of a mass reduction according to Brown-Rho scaling~\cite{Brown} and an
additional increase in the decay width due to collisional broadening of
$\Gamma_{coll}\approx$50~MeV,
a shift in the $\omega$ mass can still be experimentally observed.
In particular, differences in the $\pi^0\gamma$ invariant-mass spectrum
for several nuclei might provide evidence for a mass shift.

The dominant experimental background is expected to stem from the process
$\gamma+A\rightarrow \pi^0\pi^0+X$. By missing one of the photons,
this channel can be misidentified as a $\gamma+A\rightarrow \pi^0\gamma+X$
reaction. This background has been estimated using the coupled-channel model, and
it is found to be small at the $\pi^0\gamma$-mass region where a mass shift of
the $\omega$ meson is expected. Kinematic cuts on the energy of the photons
have demonstrated to be an efficient technique to reduce this source of contamination.

%Conclusion.....

In conclusion, it is shown that the $\gamma+A\rightarrow\pi^0\gamma$ process
can be used experimentally to measure the in-medium mass of the $\omega$ meson.
This conclusion is based upon assumptions on properties of the in-medium
$\omega$ meson, like mass and width, which theoretically are considered to
be realistic.

\section*{Acknowledgements}

We gratefully acknowlegde discussions with the CB-ELSA group,
in particular with U.~Thoma. This work is supported by DFG and FZ
J\"ulich.

\end{document}